\documentclass[pdftex,10pt,twocolumn]{revtex4}
\usepackage{color,afterpage}
\usepackage{listings}
\usepackage{hyperref}
\hypersetup{ pdfborder=0,
 breaklinks,
 baseurl=http://,
 pdfpagemode=None,
 pdfstartview=XYZ,
 pdfstartpage=1,
 pdftitle    = {A statistical method to identify water masses},
 pdfsubject  = {The oceanographic time series in Fram Strait},
 pdfauthor   = {fgreil<at>awi.de},
 pdfkeywords = {Alfred Wegener Institute for Polar- and Marine Research},
 pdfcreator  = {Adobe-Acrobat-Distiller},
 pdfproducer = {R, vim, LaTeX with hyperref}
 bookmarksopen=false,   
}

\usepackage[dvips]{graphicx}

\usepackage{amsmath,amssymb,psfrag}
\usepackage{longtable}

\let\oldmarginpar\marginpar
\renewcommand\marginpar[1]{\-\oldmarginpar[\raggedleft\tiny #1]%
{\raggedright\tiny #1}}

\begin{document} 
\title{A new method to identify water masses -- a network-based\\
analysis of oceanographic point measurement time series}
\author{Florian Greil \href{mailto:florian.greil@awi.de}{\tt
<fgreil@awi.de>}}
\affiliation{Stiftung Alfred-Wegener-Institut f\"{u}r Polar- und
Meeresforschung in der Helmholtz-Gemeinschaft,
Am Handelshafen 12, 27570~Bremerhaven, Germany}
\date{\today}
\begin{abstract}
This is a statistical analysis of the oceanographic time series measured
across Fram Strait at a latitude of $78^\circ50'$\,N. Fram Strait is the
deepest passage between the Arctic Ocean and the North Atlantic. There are
up to 16\,mooring lines with instruments at different depths measuring
water temperature and velocity. These variables vary on different time
scales and the challenge is to distinguish different spatial flow regimes.

For Fram Strait, a temperature criterion is traditionally applied to identify
\emph{water-masses}, i.e.\,water volumes of similar origin.  Interpolation
leads to a vertical latitudinal 2D cross-section from which a scalar -- the
hypothetical area of waters within a certain temperature interval -- can
be extracted. The scalar is combined with a similar interpolation of the
velocities to approximate the volume flows through the gateway. This
approach is not only numerically expensive but also incorporates many
assumptions.
The present study suggest a new network-based approach to discriminate
between flow regimes without the need to introduce artificial data through
interpolation.  The new approach not only reproduces the known flow
patterns, but also reveals topographical features which are not captured by
a standard water-masses analysis. 
\end{abstract} \typeout{FG> End abstract ------}
\maketitle 

Network theory offers a appealing framework to study local and non-local
relationships of spatio-temporal variables. Global climate networks are
reconstructed from fields of observables such as ocean temperature at sea
level \cite{Donges}. Network analysis of the temperature field uncovered
wave-like structures of energy flow related to surface ocean currents.
Having an unique set of highly resolved oceanographic time series in hand,
network \mbox{analysis} is here used to gain insight into the flow patterns
at Fram Strait. The data set is the first network which ``goes into the
deep'' of the ocean, i.e.\,where the network is laterally extended.

\begin{figure*}[t!]
\includegraphics[width=\textwidth]{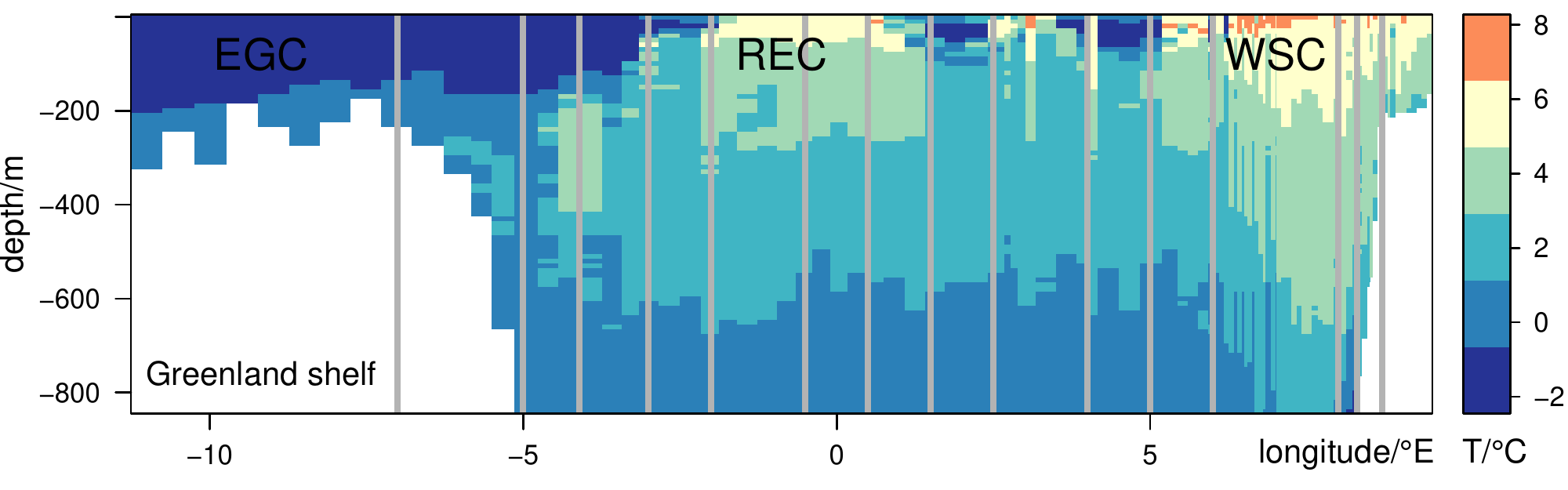}\\
\vspace{-0.6cm}
\caption{Temperature field of the upper 800\,m of Fram Strait at a latitude
of 78\,$^\circ$50'\,N. The data was acquired during a cruise with
RV~Polastern between June~24th and July~11th, 2011. White areas correspond
to sea floor. The bold perpendicular lines symbolize the longitudes of the
mooring lines, a degree longitude corresponds to roughly 20\,km here.}
\label{FigCtd}
\end{figure*}

Fram Strait is approx.\,300\,km wide in the deep part, considering the
(shallow) continental shelf leads to a width of 500\,km. Therefore, the
current study is one of the first \emph{regional} climate networks
\cite{WoonSeon}.  The data basis here is an experimental measurement series
instead of reanalysis data. The advantage is that every feature found by
the network analysis can directly be judged by the experienced
oceanographer. A yearly verification of the inter- and extrapolation is
offered by a spatially high resolved hydrographic survey.  The survey is
carried during each cruise for deploying or recovering the moored
instruments.  It delivers a quasi-instantaneous snapshot of the water
properties across Fram Strait (Fig.\,\ref{FigCtd} shows the situation in
2011).  In the following, the under-water buoys will be referred as
\emph{moorings}.

\begin{figure}[t!]
\includegraphics[width=\columnwidth]{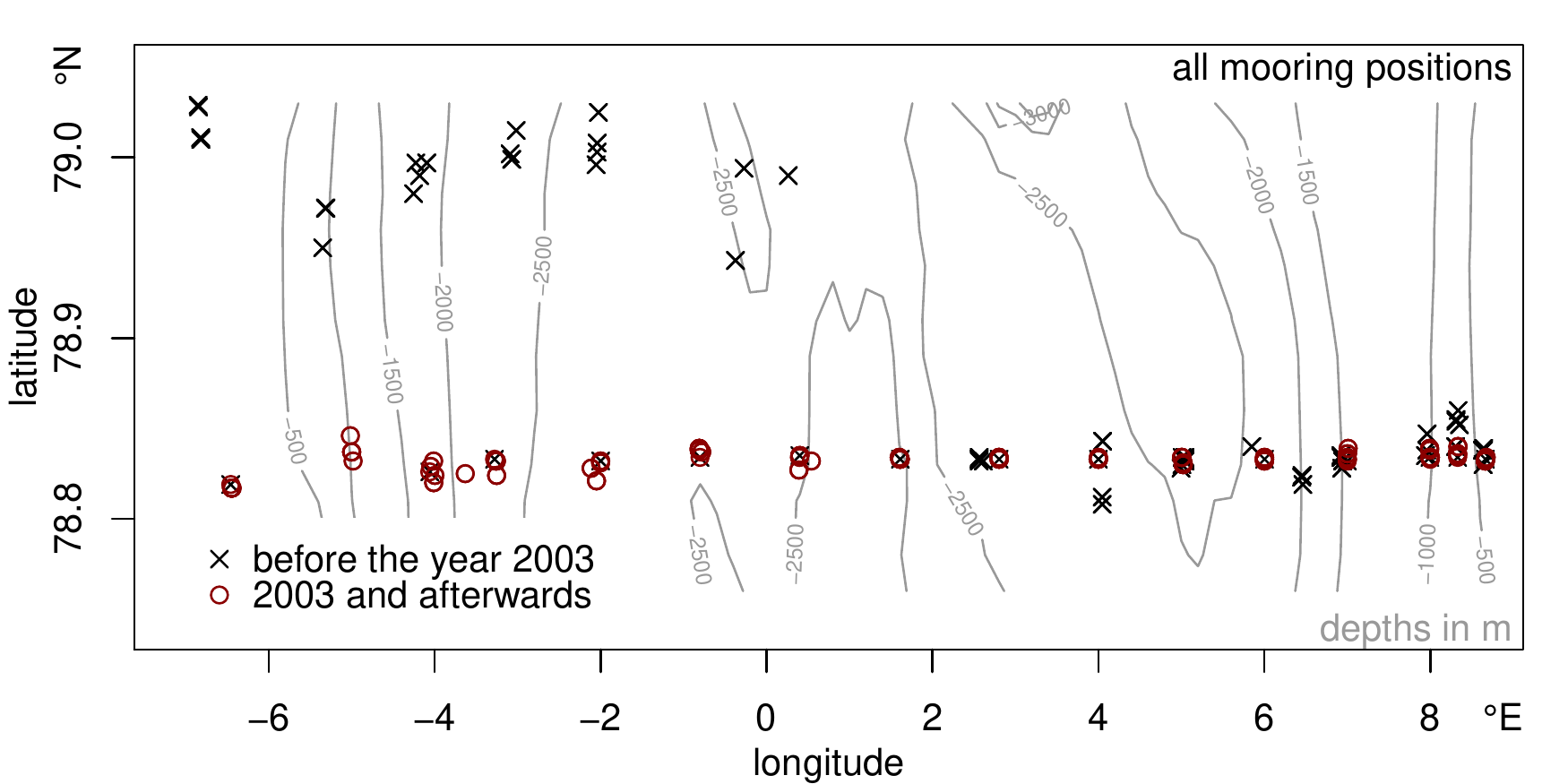}
\vspace{-0.6cm}
\caption{Map of the mooring positions in Fram Strait. The meridional distance between
78.8$^\circ$N and 79$^\circ$N is around 20\,km which is the same as the
zonal spacing. For our analysis only the measurements from 2003 and
afterward ($\circ$) are taken.} \label{FigBathy} 
\end{figure}

\section{The Arctic Ocean}
The Arctic Ocean is subject to changes e.g.\,as shrinking sea ice
extent and the additional thinning of the ice. Warm impulses of water from 
the Atlantic Ocean are circulating in the Arctic basins. 
Water from the Pacific Ocean transports internal energy (``oceanic
heat'') through Bering Strait into the Arctic Ocean. Changes of sea ice
extent, thickness and volume are observed by remote
sensing and by drifting buoys with measuring equipment inside
\cite{Perovich}.  Monitoring the state of the water below the ice is
technically difficult and requires international efforts in terms of
oceanographic surveys with ice-breakers or multiyear observatories deployed
at the seafloor.

The approach to gain synoptic (i.e.\,large-scale) insights about the Arctic
Ocean exchanges is to monitor the in- and out-flow of water through Fram
Strait. Although climatological scales are defined over periods of 30~years
and more according to the World Meteorological Organization
(\href{http://wmo.int}{WMO}), it seems possible to set up a preliminary
energy- and volume balances: There are only four key gateways to the Arctic
Ocean where water and sea-ice leaves or enters \cite{ChangingArctic}. The
Fram Strait is the only deep-sea opening of the Arctic Ocean to the other
oceans, all other gateways are shallower and/or narrower, see
Tab.\,\ref{TabGateways}. 

\begin{table}[b]
\begin{tabular*}{\columnwidth}{@{\extracolsep\fill}p{3cm}ccc}
\hline
Arctic gateway    & width & max.\,depth& net volume flux \\
\hline \hline
Barents Sea Opening&350\,km& 500\,m& $2.0\pm2.9$\,Sv \\
\hline
Fram Strait   & 300\,km &2\,600\,m& $-2.0\pm2.7$\,Sv\\
\hline
Bering Strait   & 86\,km  & 50\,m & $0.8\pm 0.2$\,Sv \\
\hline
Nares Strait    & 40\,km  &220\,m &$0.57\pm 0.3$\,Sv\\
Lancaster Sound & 40\,km  &125\,m &$-0.7$\,Sv \\
Cardigan Strait & 8\,km   &180\,m &$-0.3$\,Sv \\
\hline
\end{tabular*} ~\vspace{-0.3cm}%
\caption{Gateways to the Arctic Ocean, the last three are the largest 
of the Canadian Archipelago. 1~Sverdrup (Sv) = $10^6 {\rm m}^3/{\rm s}$.
Emptying Lake Constance (German: ``Bodensee'') within a day corresponds to
0.56\,Sv. Estimates from \cite{ChangingArctic}, positive values: 
inflow into the Arctic Ocean.} \label{TabGateways}
\end{table}

\section{The Fram Strait time series}
The two main currents of Fram Strait are a northward inflow of water from
the Atlantic (next to Spitsbergen) on the one hand and a southward outflow
of polar freshwater (next to Greenland) on the other hand.
The East Greenland Current (EGC) can be recognized as (dark blue) cold patch
above 200\,m at the left side of Fig.\,\ref{FigCtd}. The water there moves
southwards (out of the paper plane) and carries sea ice out of the Arctic
Ocean.  The \emph{West Spitsbergen Current} (WSC) is the patch with
temperatures above $2^\circ$\,C (red, yellow, light-green) on the right side
of Fig.\,\ref{FigCtd}, east of $5^\circ$\,E. The WSC is a branch of the
North Atlantic Current which is the northern extension of the Gulf Stream.
The WSC is the warmest water mass entering the Arctic Ocean. Since our
measurements started in 1997, the temperature of the Atlantic water
increased with a rate of around $1^\circ$\,C within 10\,years
\cite{ChangingArctic}.

Motivated by the importance of the Fram Strait for the Arctic climate, the
Alfred-Wegener-Institute for Polar- and Marine Research (AWI) maintains a
transect of moorings in collaboration with the Norwegian Polar Institute at
$78^\circ50'$\,N. The transect ends at the shelf breaks at $6^\circ 52$W
and $8^\circ40$E respectively, see Fig.~\ref{FigCtd} and
Fig.~\ref{FigBathy}.

Instruments (from the companies Aanderaa and Falmouth Scientific
Instruments) provide point measurements of horizontal velocities, flow
direction, temperature and salinity. However, the salinity measurements of
the current meters are not trustworthy since the conductivity cell is not
pumped. This causes much slower response times than what is needed to cope
with the high flow speeds in Fram Strait. Besides, the sensors fail
frequently due to growing bio-films. Therefore, many gaps within the
salinity time series exist. The additionally available data from a few
upward-looking Acoustic Doppler current profilers (ADCP) near the surface
are also neglected as their time-series are far shorter.

\begin{figure*}[htbp]
\includegraphics[width=0.91\textwidth]{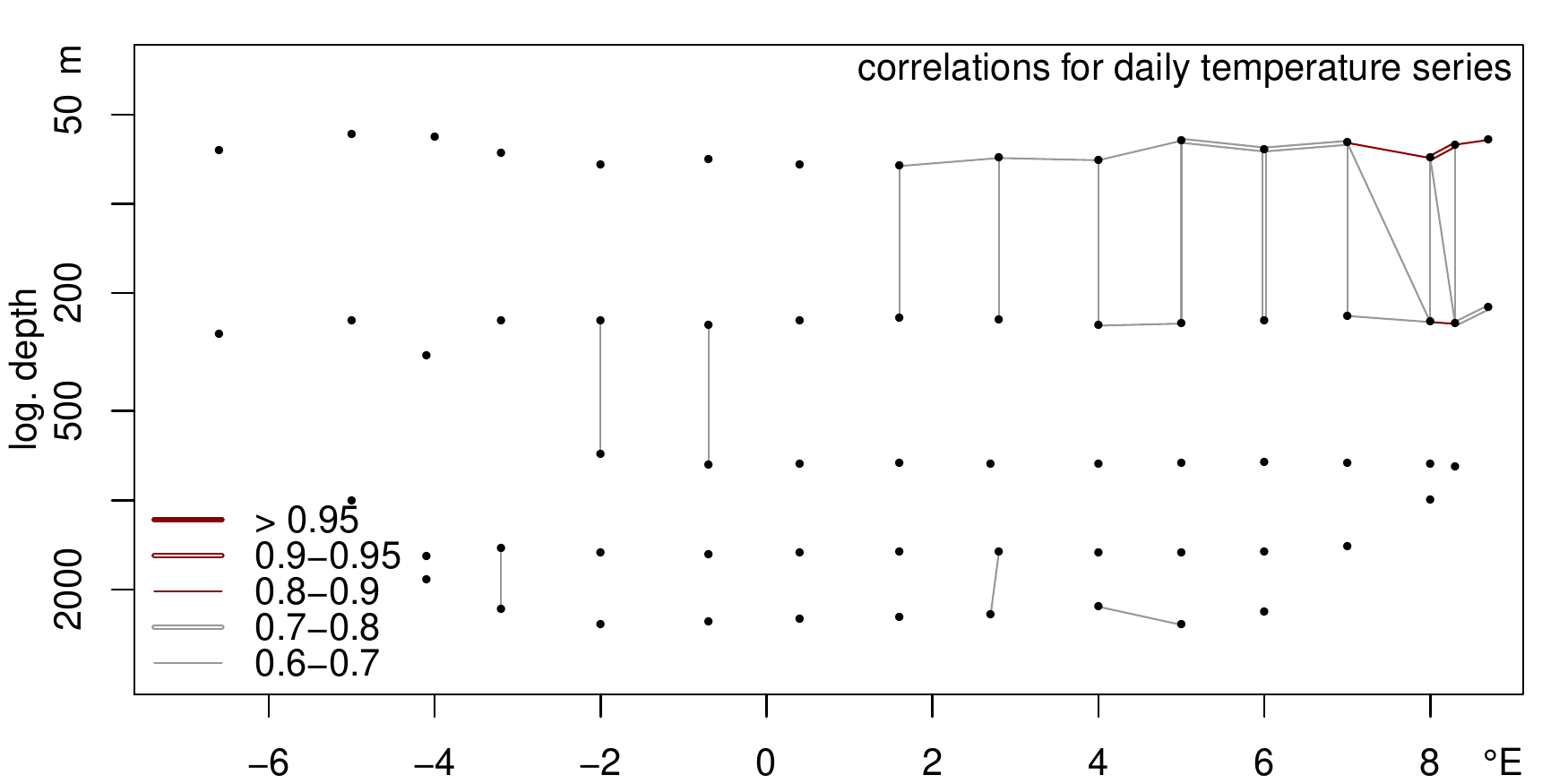}\\
\vspace{-1cm}%
\resizebox{\textwidth}{1cm}{\textcolor{white}{\rule{1in}{2in}}}\\
\vspace{-1.5cm}
\includegraphics[width=0.91\textwidth]{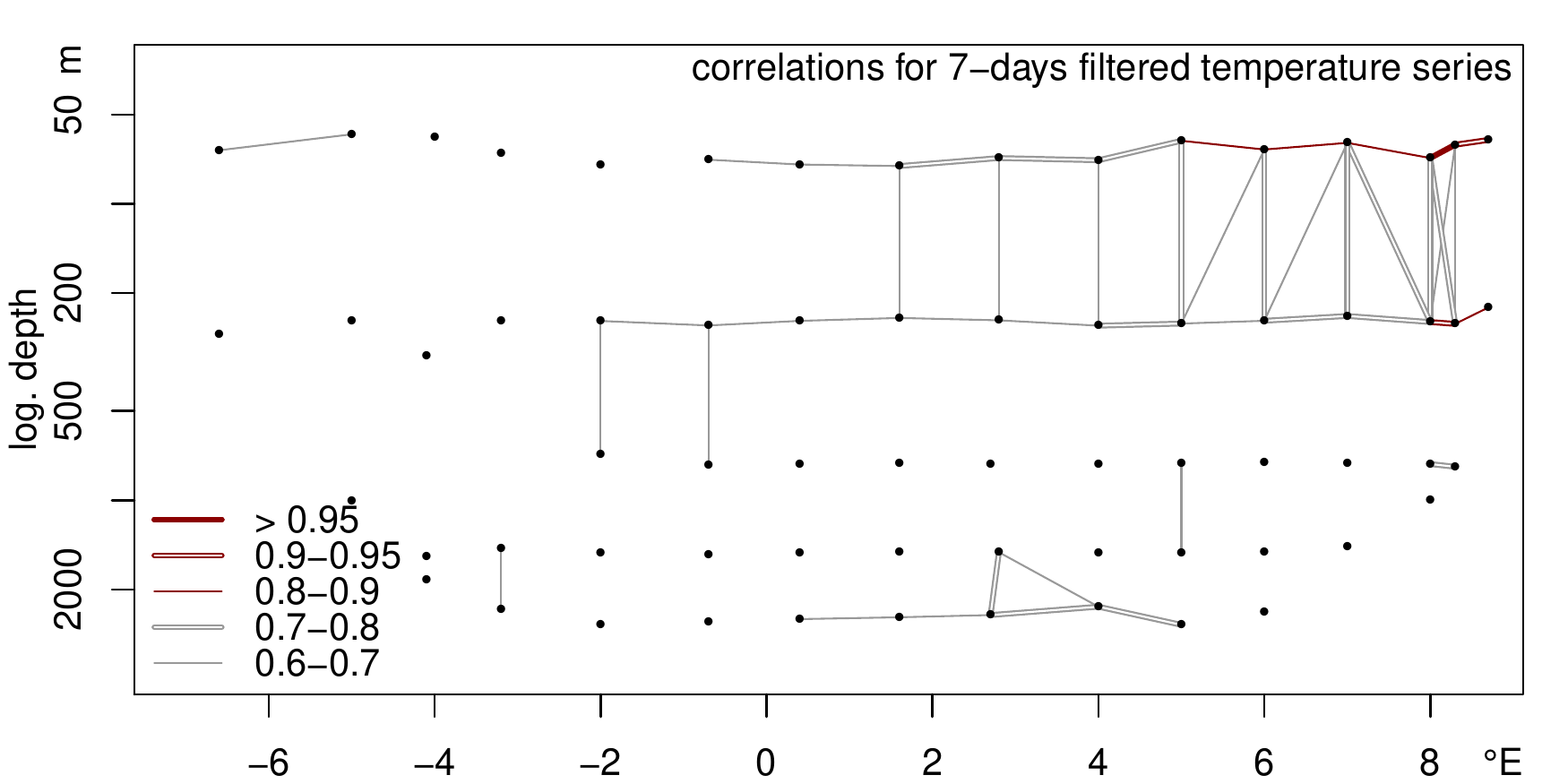}\\
\vspace{-1cm}%
\resizebox{\textwidth}{1cm}{\textcolor{white}{\rule{1in}{2in}}}\\
\vspace{-1.5cm} 
\includegraphics[width=0.91\textwidth]{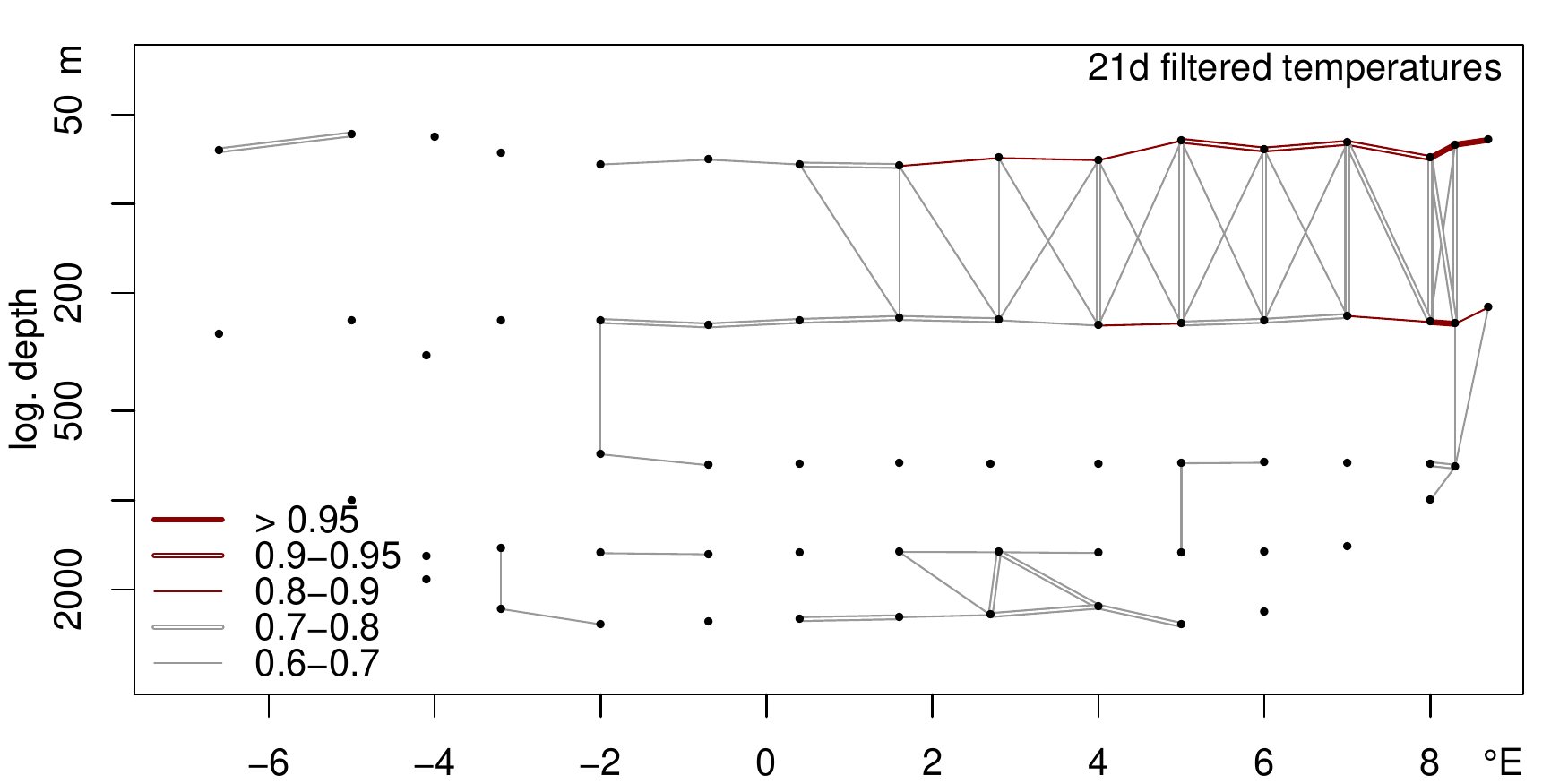}\\
\vspace{-0.4cm}
\caption{Temperature correlations coefficients between nearest and 2nd
nearest neighbors of all timeseries since 2003 for daily means (top) and
with a 7-day filter (bottom). The bullets represent the average instrument
position (depth and longitude), for this reason the mooring lines are not
straight as sketched in Fig.\ref{FigCtd}.} \label{FigTemp} 
\end{figure*}

Data is recorded at 2 to 5 different depth levels every hour.  The lowest
level usually is located approximately 10\,m above the sea-floor. If
applicable, there are instruments at 1500\,m depth, around 800\,m and
300\,m depth. The uppermost sensor is  located around 60\,m below the sea
surface to avoid losses of instruments by drifting ice keels. Recently,
fish trawling became more frequent with the retreating ice edge. The
fishing nets are a threat for our instruments.

In-situ data by observational oceanography are assimilated by ocean
modelers. A major product for them are time series of de-tided daily
averages. The effect of tides are excluded by filtering out semi-diurnal
and diurnal tidal constituents. The analysis is based on the daily means,
accepting that any additional post-processing of the measurements
(e.g.\,depth-correction to due flow-induced dives or manual spike-removal)
happens inside a black box.  For our purposes these are only minor
changes which do not change the main statements.

\afterpage{\clearpage}

\section{Constructing a correlation network}
The method suggested here should work for any set of spatio-temporal time
series. In Fram Strait, the network of point measurements has 66~nodes with
three times series per node~$x$: the temperature~$T_x$, the meridional
velocity~$v_x$ and the zonal velocity~$u_x$ at a given position~$x \in
\{1,\ldots,66\}$. Each $x$ can be mapped on a longitudinal position and a
depth level. Every time series is constructed by merging snippets of the
maintenance period.

\begin{figure*}[t!]
\includegraphics[width=0.91\textwidth]{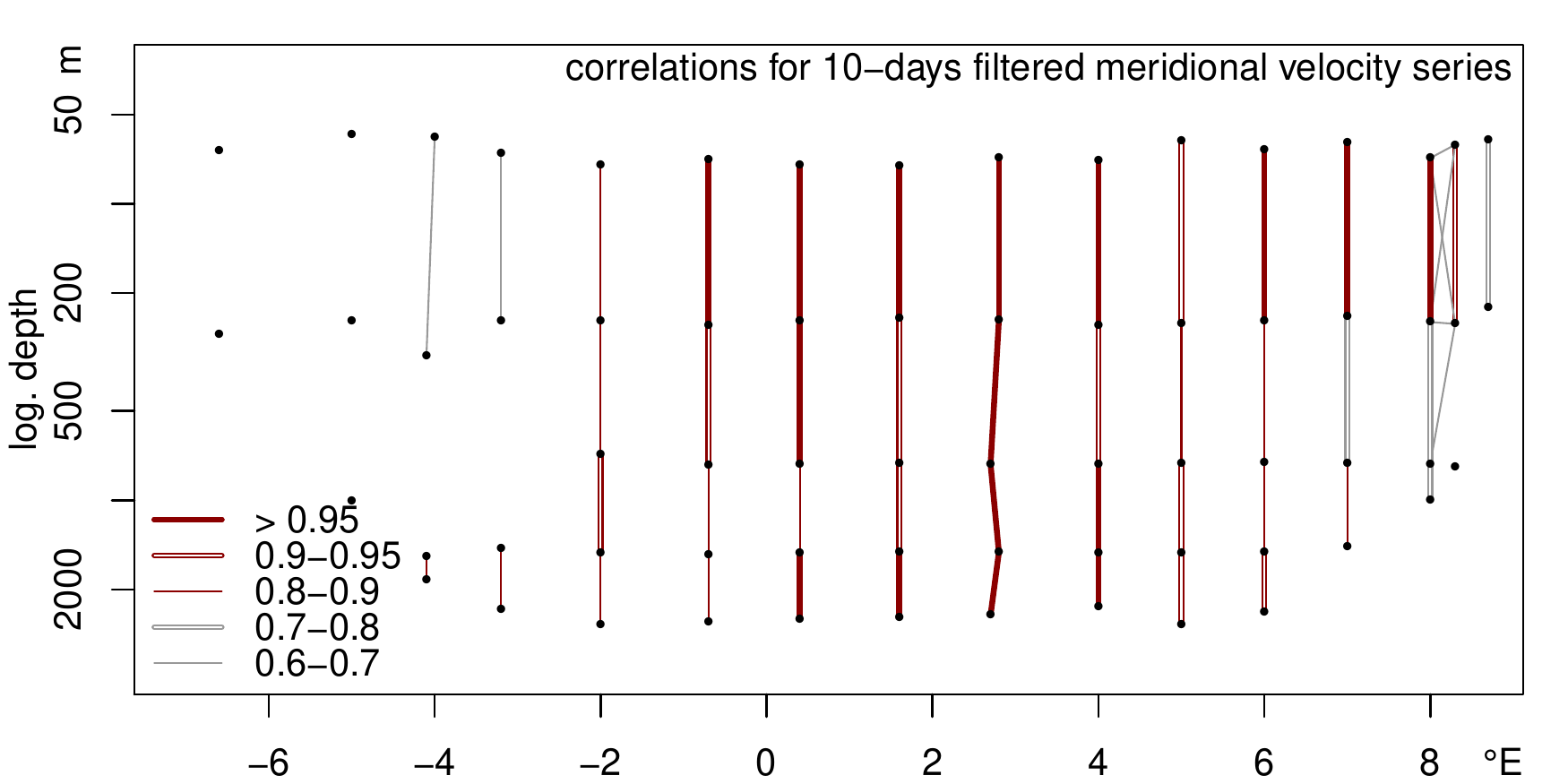}\\
\vspace{-1cm}%
\resizebox{\textwidth}{1cm}{\textcolor{white}{\rule{1in}{2in}}}\\
\vspace{-1.5cm} 
\includegraphics[width=0.91\textwidth]{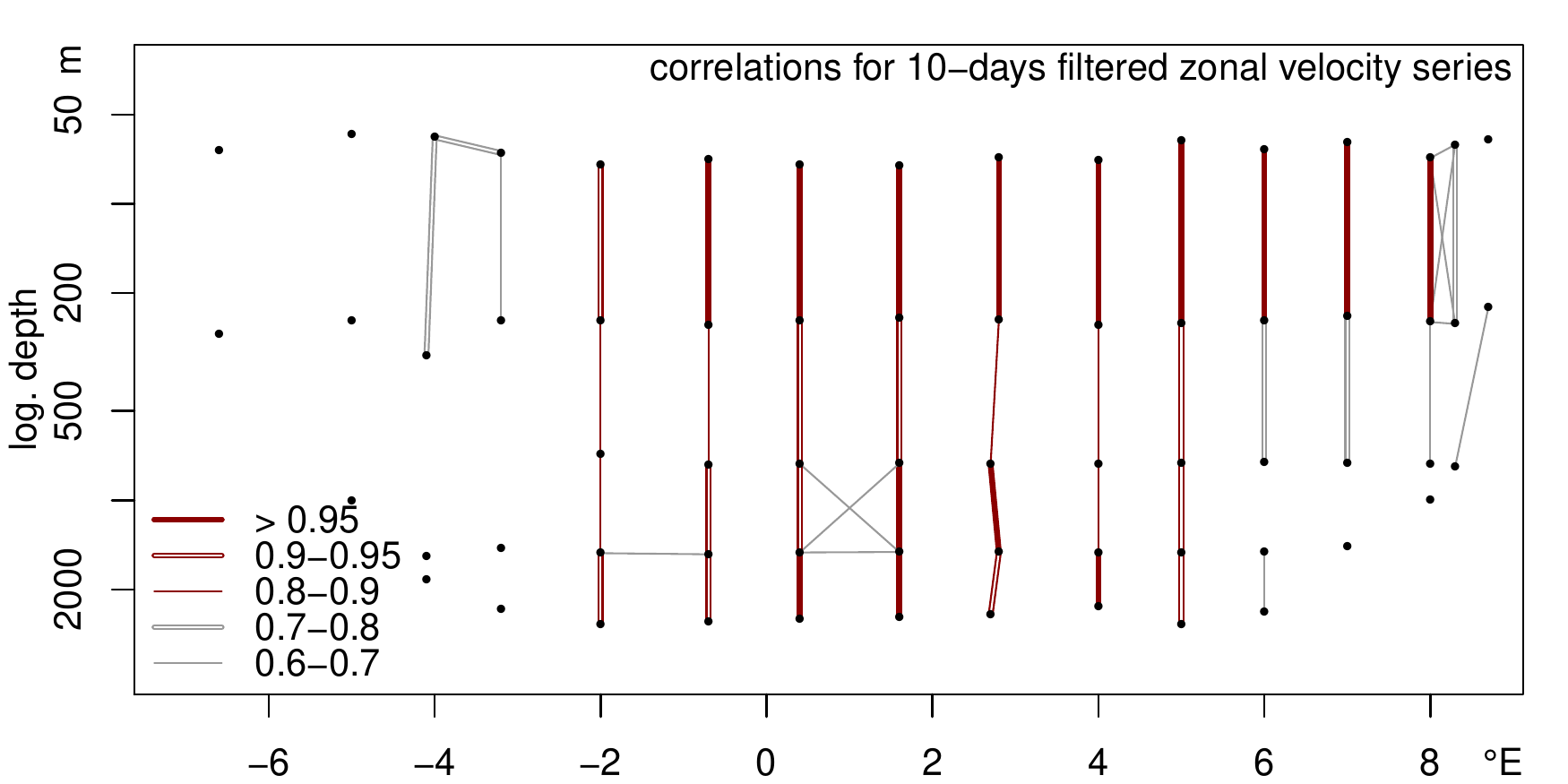}\\
\vspace{-0.4cm}
\caption{Correlation graph for meridional (top) and zonal (bottom)
velocities with a 10-day filter for the complete time series since 2003.}
\label{FigVel}
\end{figure*}

The Spearman correlation coefficient~$\rho$ between any nearest or
second-nearest neighbors is calculated for the complete time series since
2003. Working with daily averages, only correlations between observables at
the same time instance are considered. Pairwise complete observables are
taken into account, i.e.\,whenever both time series have a valid value at a
given point in time.  Looking at \emph{all} different pairs of correlation
is a standard consistency check and therefore $66^2=4365$ correlation
coefficients have to be evaluated. New is the concise and intuitive
representation as a network of (2nd)~nearest neighbors which add up to only
205~correlation coefficients.

One expects high correlations even between non-neighboring pairs of point
measurements because the dynamics of the water is driven by the atmosphere
and ocean upstream. 
The new method is to use a multi-scale approach which does not look on
the value at the correlation itself, but at how the correlation increases
with increasing filtering size.
Before, it has to be noted that correlations between neighboring
instruments are high in the vertical and significantly lower in the
horizontal direction, simply because of the different resolutions in both
directions. Within a 2\,km mooring line there are up to five instruments
while there is only a mooring line roughly every 20\,km.  The higher
vertical correlations (between velocities within same mooring line) are
usually interpreted as sign of a prevailing barotropic structure of the
flow, i.e.\,the flow is mainly directed by the topography of the
ground and does not depend on the depth.

Moving towards longer filtering times, the zonal correlation within the
(sub)surface widens towards the West. This is expected when the upper
ocean is well-mixed by the atmosphere. In particular, where the ocean
surface is only rarely covered by sea ice, mixing takes place. The strong
correlations at the bottom between $2$ and $4^\circ$\,East are most
probably induced by the topographic steering of the Knippovich Ridge whose
Northern foothill mountains are located at these longitudes,
see Fig.\,\ref{FigBathy}.  Besides, there are vertical correlations at the
correct depth and longitudinal position where the Atlantic Water is
expected to recirculate. The recirculation manifests itself in form of
coherent southward flow. A hint is the warm patch ($T \geq
4^\circ$C) between $\pm 3^\circ$E in Fig.\,\ref{FigCtd} (REC). The
long-term averaged velocity field (not shown here) supports this
assumption.

The effect of the filtering time on the pattern is small for the
correlation network constructed from the water velocities. Therefore,
without loss of generality a 10-days boxcar filter is applied in
Fig.\,\ref{FigVel}.

\section{Multiscale correlation analysis.}
In oceanography, a water mass is ``an identifiable water volume with a
common history which may be traced back to some source area''.
Traditionally, water masses are defined by (potential) temperature and
salinity, although additional properties may also taken into account.
Ideally, these are conservative, in the sense that they are only modified
by mixing, and neither biological activity, nor chemical degradation
influences the parameter. In Fram Strait, no reliable salinity measurements
for the whole time series exist. Thus, a pure temperature criterion has
been introduced to identify different water masses. Whenever the
temperature is above a threshold of $1^\circ$C the water is assumed to
originate in the Atlantic \cite{FramSeriesA}. Here, a statistical method to
identify water masses is presented. This approach can be also helpful for
other networks of time series.

The correlation pattern between the 2nd and 3rd mooring line from the East
is independent of the variable, see Fig.\,\ref{FigTemp} and
\ref{FigVel}. It is exactly the location of the central core of the West
Spitsbergen Current (WSC) which is consistent with the
temperature-criterion for identifying the water masses, compare
Fig.\,\ref{FigCtd}.

It is even more interesting to identify regions where correlations are lower and
the absence of links persists when changing the filtering time. All
correlations grow slowly with increasing filtering intervals since the
filtering operation introduces correlations. However, the effect of growing
filter window also provides oceanographic information. For instance the
missing vertical links in in the Western part of the Strait (for all
variables, but more pronounced for the temperature-correlation network) are
a clear sign for the decoupling of the ocean from the atmosphere which
occurs due to frequent sea ice cover in this region. For the water
velocities in Fig.\,\ref{FigVel}, the otherwise prominent vertical
correlations disappear between the 2nd and 3rd layer of instruments. Exactly
in those regions the two boundary currents tend to have their vertical
limits.

\section{Conclusion}
Traditional methods to classify flow regimes rely on numerical expensive
interpolation schemes. The drawback is that values (for temperature and
water velocity) have to produced for the complete vertical field across the
gateway. The interpolation is based on sparse measurements every 20\,km
which necessarily comes along with large error bars for transport
estimates, see \cite{FemSect} for a reliable method based on finite
elements. 

A new climate network has been introduced which allows to identify water
masses and other spatially consistent flow regimes using simple maths.  The
nodes are the instrument positions, given as longitude-depth coordinates.
Links are established if the correlation between the time series between
two nodes is bigger than a certain threshold. 

The new approach allows to quantify regions of interest without any prior
knowledge about oceanographic phenomena. Where links are absent within the 
correlation network, increasing the spatial resolution of the transect may
help to gain precision. One crucial region is the longitudinal resolution
at 6$\pm1^\circ$\,E. The new method quantifies the common knowledge that the
error of the flux estimates actually originates in the uncertainty of the
western boundary of the West Spitsbergen Current (WSC). 

\section{Data sources} \label{SecDataSources}

Eventually the complete data set will be freely available, the following
table summarizes the state in June 2012.
{\tiny 
\begin{longtable}{p{1cm}p{1.55cm}p{1.55cm}p{0.85cm}p{0.85cm}p{2.2cm}}
\hline
snippet  & begin & end & $~^\circ W$ & $~^\circ N$ & doi:10.1594/ \\
\hline
F1-7 & 2004-07-20 & 2005-08-17 & 8.664 & 78.832 & 
\href{http://dx.doi.org/10.1594/pangaea.778856}{pangaea.778856}\\
F1-8 & 2005-08-17 & 2006-08-21 & 8.664 & 78.832 & 
\href{http://dx.doi.org/10.1594/pangaea.778857}{pangaea.778857}\\
F1-10 & 2007-09-12 & 2008-07-05 & 8.674 & 78.834 & 
\href{http://dx.doi.org/10.1594/pangaea.777564}{pangaea.777564}\\ 
F1-11 & 2008-07-07 & 2009-07-04 & 8.667 & 78.833 & 
\href{http://dx.doi.org/10.1594/pangaea.777565}{pangaea.777565}\\
F2-6 & 2002-08-02 & 2003-09-22 & 8.330 & 78.834 & 
\href{http://dx.doi.org/10.1594/pangaea.780554}{pangaea.780554}\\
F2-8 & 2004-07-20 & 2005-08-17 & 8.327 & 78.836 & 
\href{http://dx.doi.org/10.1594/pangaea.778867}{pangaea.778867}\\
F2-9 & 2005-08-18 & 2006-08-21 & 8.327 & 78.836 & 
\href{http://dx.doi.org/10.1594/pangaea.778868}{pangaea.778868}\\
F2-10 & 2006-08-23 & 2007-09-11 & 8.327 & 78.836 & 
\href{http://dx.doi.org/10.1594/pangaea.777574}{pangaea.777574}\\
F2-11 & 2007-09-28 & 2008-07-05 & 8.329 & 78.835 & 
\href{http://dx.doi.org/10.1594/pangaea.777575}{pangaea.777575}\\
F2-12 & 2008-07-07 & 2009-07-04 & 8.333 & 78.840 & 
\href{http://dx.doi.org/10.1594/pangaea.777576}{pangaea.777576}\\
F2-13 & 2009-07-05 & 2010-07-01 & 8.334 & 78.840 & 
\href{http://dx.doi.org/10.1594/pangaea.777577}{pangaea.777577}\\
F2-14 & 2010-07-03 & 2011-06-25 & 8.334 & 78.834 & 
\href{http://dx.doi.org/10.1594/pangaea.777578}{pangaea.777578}\\
F3-6 & 2003-09-26 & 2004-07-19 & 7.994 & 78.834 & 
\href{http://dx.doi.org/10.1594/pangaea.778908}{pangaea.778908}\\
F3-7 & 2004-07-20 & 2005-08-17 & 7.992 & 78.838 & 
\href{http://dx.doi.org/10.1594/pangaea.778869}{pangaea.778869}\\
F3-8 & 2005-08-18 & 2006-08-22 & 7.992 & 78.839 & 
\href{http://dx.doi.org/10.1594/pangaea.778870}{pangaea.778870}\\
F3-9 & 2006-08-23 & 2007-09-11 & 7.992 & 78.839 & 
\href{http://dx.doi.org/10.1594/pangaea.777583}{pangaea.777583}\\
F3-10 & 2007-09-28 & 2008-07-05 & 8.001 & 78.834 & 
\href{http://dx.doi.org/10.1594/pangaea.777579}{pangaea.777579}\\
F3-11 & 2008-07-07 & 2009-07-03 & 8.000 & 78.833 & 
\href{http://dx.doi.org/10.1594/pangaea.777580}{pangaea.777580}\\
F3-12 & 2009-07-05 & 2010-07-01 & 8.001 & 78.834 & 
\href{http://dx.doi.org/10.1594/pangaea.777581}{pangaea.777581}\\
F3-13 & 2010-07-03 & 2011-06-25 & 8.000 & 78.833 & 
\href{http://dx.doi.org/10.1594/pangaea.777582}{pangaea.777582}\\
F4-6 & 2003-09-26 & 2004-07-18 & 7.000 & 78.833 & 
\href{http://dx.doi.org/10.1594/pangaea.778871}{pangaea.778871}\\
F4-7 & 2004-07-18 & 2005-08-17 & 7.000 & 78.836 & 
\href{http://dx.doi.org/10.1594/pangaea.778872}{pangaea.778872}\\
F4-8 & 2005-08-18 & 2006-08-23 & 7.002 & 78.836 & 
\href{http://dx.doi.org/10.1594/pangaea.778873}{pangaea.778873}\\
F4-9 & 2006-08-27 & 2007-09-16 & 7.010 & 78.839 & 
\href{http://dx.doi.org/10.1594/pangaea.777588}{pangaea.777588}\\
F4-10 & 2007-09-12 & 2008-07-05 & 6.997 & 78.834 & 
\href{http://dx.doi.org/10.1594/pangaea.777584}{pangaea.777584}\\
F4-11 & 2008-07-07 & 2009-07-03 & 7.000 & 78.833 & 
\href{http://dx.doi.org/10.1594/pangaea.777585}{pangaea.777585}\\
F4-12 & 2009-07-04 & 2010-07-01 & 7.000 & 78.833 & 
\href{http://dx.doi.org/10.1594/pangaea.777586}{pangaea.777586}\\
F4-13 & 2010-07-04 & 2011-06-19 & 7.006 & 78.835 & 
\href{http://dx.doi.org/10.1594/pangaea.777587}{pangaea.777587}\\
F5-6 & 2003-10-02 & 2004-07-17 & 6.002 & 78.832 & 
\href{http://dx.doi.org/10.1594/pangaea.778874}{pangaea.778874}\\
F5-7 & 2004-07-19 & 2005-08-18 & 6.000 & 78.832 & 
\href{http://dx.doi.org/10.1594/pangaea.778875}{pangaea.778875}\\
F5-8 & 2005-08-23 & 2006-08-28 & 6.003 & 78.833 & 
\href{http://dx.doi.org/10.1594/pangaea.778876}{pangaea.778876}\\
F5-9 & 2006-08-28 & 2007-09-11 & 6.003 & 78.834 & 
\href{http://dx.doi.org/10.1594/pangaea.777593}{pangaea.777593}\\
F5-10 & 2007-09-12 & 2008-07-05 & 6.000 & 78.834 & 
\href{http://dx.doi.org/10.1594/pangaea.777589}{pangaea.777589}\\
F5-11 & 2008-07-12 & 2009-07-03 & 6.000 & 78.833 & 
\href{http://dx.doi.org/10.1594/pangaea.777590}{pangaea.777590}\\
F5-12 & 2009-07-05 & 2010-07-01 & 6.000 & 78.833 & 
\href{http://dx.doi.org/10.1594/pangaea.777591}{pangaea.777591}\\
F5-13 & 2010-07-04 & 2011-06-25 & 6.000 & 78.833 & 
\href{http://dx.doi.org/10.1594/pangaea.777592}{pangaea.777592}\\
F6-7 & 2003-09-27 & 2004-07-17 & 5.021 & 78.830 & 
\href{http://dx.doi.org/10.1594/pangaea.778877}{pangaea.778877}\\
F6-8 & 2004-07-19 & 2005-08-23 & 5.022 & 78.830 & 
\href{http://dx.doi.org/10.1594/pangaea.778878}{pangaea.778878}\\
F6-9 & 2005-08-26 & 2006-08-25 & 5.022 & 78.830 & 
\href{http://dx.doi.org/10.1594/pangaea.778879}{pangaea.778879}\\
F6-10 & 2006-08-28 & 2007-09-11 & 5.022 & 78.831 & 
\href{http://dx.doi.org/10.1594/pangaea.777594}{pangaea.777594}\\
F6-11 & 2007-09-13 & 2008-07-11 & 5.002 & 78.834 & 
\href{http://dx.doi.org/10.1594/pangaea.777595}{pangaea.777595}\\
F6-12 & 2008-07-12 & 2009-07-03 & 5.004 & 78.834 & 
\href{http://dx.doi.org/10.1594/pangaea.777596}{pangaea.777596}\\
F6-13 & 2009-07-06 & 2010-07-02 & 5.004 & 78.834 & 
\href{http://dx.doi.org/10.1594/pangaea.777597}{pangaea.777597}\\
F6-14 & 2010-07-02 & 2011-06-27 & 5.000 & 78.834 & 
\href{http://dx.doi.org/10.1594/pangaea.777598}{pangaea.777598}\\
F7-6 & 2004-07-22 & 2005-08-23 & 4.000 & 78.833 & 
\href{http://dx.doi.org/10.1594/pangaea.778880}{pangaea.778880}\\
F7-7 & 2005-08-26 & 2006-08-24 & 4.000 & 78.833 & 
\href{http://dx.doi.org/10.1594/pangaea.778881}{pangaea.778881}\\
F7-8 & 2006-08-29 & 2008-07-11 & 4.000 & 78.834 & 
\href{http://dx.doi.org/10.1594/pangaea.777600}{pangaea.777600}\\
F7-9 & 2008-07-15 & 2010-07-10 & 3.997 & 78.833 & 
\href{http://dx.doi.org/10.1594/pangaea.777601}{pangaea.777601}\\
F7-10 & 2010-07-11 & 2011-06-27 & 4.000 & 78.833 & 
\href{http://dx.doi.org/10.1594/pangaea.777599}{pangaea.777599}\\
F8-7 & 2004-07-22 & 2005-08-29 & 2.801 & 78.834 & 
\href{http://dx.doi.org/10.1594/pangaea.778882}{pangaea.778882}\\
F8-8 & 2005-08-31 & 2006-08-29 & 2.802 & 78.834 & 
\href{http://dx.doi.org/10.1594/pangaea.778883}{pangaea.778883}\\
F8-9 & 2006-08-29 & 2008-07-15 & 2.801 & 78.833 & 
\href{http://dx.doi.org/10.1594/pangaea.777604}{pangaea.777604}\\
F8-10 & 2008-07-18 & 2010-07-10 & 2.805 & 78.833 & 
\href{http://dx.doi.org/10.1594/pangaea.777602}{pangaea.777602}\\
F8-11 & 2010-07-11 & 2011-06-27 & 2.799 & 78.833 & 
\href{http://dx.doi.org/10.1594/pangaea.777603}{pangaea.777603}\\
F9-6 & 2004-08-21 & 2005-08-29 & -0.812 & 78.839 & 
\href{http://dx.doi.org/10.1594/pangaea.778884}{pangaea.778884}\\
F9-7 & 2005-08-30 & 2006-09-07 & -0.811 & 78.838 & 
\href{http://dx.doi.org/10.1594/pangaea.778885}{pangaea.778885}\\
F9-8 & 2006-09-08 & 2008-07-20 & -0.811 & 78.839 & 
\href{http://dx.doi.org/10.1594/pangaea.777605}{pangaea.777605}\\
F9-9 & 2008-07-21 & 2010-07-18 & -0.782 & 78.837 & 
\href{http://dx.doi.org/10.1594/pangaea.777606}{pangaea.777606}\\
F10-6 & 2003-09-30 & 2004-08-21 & -2.001 & 78.832 & 
\href{http://dx.doi.org/10.1594/pangaea.778858}{pangaea.778858}\\
F10-7 & 2004-08-24 & 2005-08-29 & -2.001 & 78.831 & 
\href{http://dx.doi.org/10.1594/pangaea.778859}{pangaea.778859}\\
F10-9 & 2006-09-09 & 2008-07-21 & -2.050 & 78.821 & 
\href{http://dx.doi.org/10.1594/pangaea.777567}{pangaea.777567}\\
F10-10 & 2008-07-21 & 2010-07-20 & -2.115 & 78.828 & 
\href{http://dx.doi.org/10.1594/pangaea.777566}{pangaea.777566}\\
F15-2 & 2003-09-28 & 2004-08-23 & 1.611 & 78.833 & 
\href{http://dx.doi.org/10.1594/pangaea.778860}{pangaea.778860}\\
F15-3 & 2004-08-23 & 2005-08-29 & 1.610 & 78.833 & 
\href{http://dx.doi.org/10.1594/pangaea.778861}{pangaea.778861}\\
F15-4 & 2005-08-30 & 2006-08-30 & 1.610 & 78.833 & 
\href{http://dx.doi.org/10.1594/pangaea.778862}{pangaea.778862}\\
F15-5 & 2006-08-30 & 2007-09-24 & 1.609 & 78.833 & 
\href{http://dx.doi.org/10.1594/pangaea.777568}{pangaea.777568}\\
F15-6 & 2007-09-24 & 2008-07-19 & 1.605 & 78.833 & 
\href{http://dx.doi.org/10.1594/pangaea.777569}{pangaea.777569}\\
F15-7 & 2008-07-18 & 2010-07-17 & 1.599 & 78.834 & 
\href{http://dx.doi.org/10.1594/pangaea.777570}{pangaea.777570}\\
F16-2 & 2003-09-29 & 2004-08-22 & 0.401 & 78.835 & 
\href{http://dx.doi.org/10.1594/pangaea.778863}{pangaea.778863}\\
F16-3 & 2004-08-22 & 2005-08-29 & 0.397 & 78.834 & 
\href{http://dx.doi.org/10.1594/pangaea.778864}{pangaea.778864}\\
F16-4 & 2005-08-30 & 2006-08-30 & 0.401 & 78.835 & 
\href{http://dx.doi.org/10.1594/pangaea.778865}{pangaea.778865}\\
F16-5 & 2006-08-31 & 2007-09-13 & 0.401 & 78.835 & 
\href{http://dx.doi.org/10.1594/pangaea.777571}{pangaea.777571}\\
F16-6 & 2007-09-13 & 2008-07-19 & 0.540 & 78.832 & 
\href{http://dx.doi.org/10.1594/pangaea.777572}{pangaea.777572}\\
F16-7 & 2008-07-20 & 2010-07-17 & 0.390 & 78.827 & 
\href{http://dx.doi.org/10.1594/pangaea.777573}{pangaea.777573}\\
\hline
\end{longtable}}
The first column is the identifier of the snippet, the first number encodes
the position, the number after the dash counts the number of deployments.

%

\end{document}